# Above Modelling and Adaptive Control of a Double Winded Induction Generator

Iosif Szeidert, Ioan Filip, Octavian Prostean, Cristian Vasar
Politehnica University of Timisoara, Department of Automation and Applied Informatics, Romania
iosif.szeidert@aut.upt.ro

*Abstract*— The paper deals with the modelling and control of a double winded induction generator. The controlled process is an induction generator with distinct excitation winding. At the generator's terminal is connected a load (electrical consumer). There are presented the results obtained by using a minimum variance adaptive control system. The main goal of the control structure is to keep the generator output (terminal voltage) constant by controlling the excitation voltage from the distinct winding. The study cases in the paper are for the validation of the reduced order model of induction generator (5$^{th}$ order model) used only to design the adaptive controller. There is also validated the control structure. There were considered variations of the mechanical torque.

## I. INTRODUCTION

The paper presents issues regarding the modelling, simulation and adaptive control of a double winded three-phased induction machine operating in generator regime. The considered induction generator has three sets of windings (for rotor, load and respectively excitation).

The model was implemented in Matlab-Simulink, there being presented study cases for the validation of the reduced order model (5$^{th}$ order) used only to design the control law. There are presented the obtained results for the minimum variance adaptive control structure. The considered control structure has as main goal to keep constant the terminal voltage of the generator under the action of an external perturbation, such as mechanical torque variation or load variation. The induction generator has a 7$^{th}$ order model that is based on the Park's equations (a nonlinear dq0 model) [1].

## II. ABOVE THE MODELLING OF THE INDUCTION GENERATOR

In figure 1, is represented the double winded induction generator in the two orthogonal dq0 axis.

The following equations (1) to (12), represent the full model of the double winded induction machine. The equations were rewritten in a convenient form in order to be implemented in dedicated simulation environment. The detailed modelling (7$^{th}$ order model) of the induction generator is presented in author's paper [1],[2],[4].

$$U_{d1} = R_1 I_{d1} + \frac{d\Psi_{d1}}{dt} - \omega_1 \Psi_{q1} \quad (1)$$

$$U_{q1} = R_1 I_{q1} + \frac{d\Psi_{q1}}{dt} + \omega_1 \Psi_{d1} \quad (2)$$

$$U_{d2} = R_2 I_{d2} + \frac{d\Psi_{d2}}{dt} - \omega_1 \Psi_{q2} \quad (3)$$

$$U_{q2} = R_2 I_{q2} + \frac{d\Psi_{q2}}{dt} + \omega_1 \Psi_{d2} \quad (4)$$

$$U_{dr} = R_3 I_{dr} + \frac{d\Psi_{dr}}{dt} - (\omega_1 - \omega)\Psi_{qr} \quad (5)$$

$$U_{dr} = R_3 I_{qr} + \frac{d\Psi_{qr}}{dt} + (\omega_1 - \omega)\Psi_{dr} \quad (6)$$

$$\Psi_{d1} = L_1 I_{d1} + M_{d12} I_{q2} + M_{d1r}(I_{dr1} - I_{qr2}) \quad (7)$$

$$\Psi_{q1} = L_1 I_{q1} + M_{q12} I_{d2} + M_{q1r}(I_{qr1} + I_{dr2}) \quad (8)$$

$$\Psi_{d2} = L_2 I_{d2} + M_{q12} I_{q1} + M_{q2r}(I_{qr1} + I_{dr2}) \quad (9)$$

$$\Psi_{q2} = L_2 I_{q2} + M_{d12} I_{d1} + M_{d2r}(I_{dr1} - I_{qr2}) \quad (10)$$

$$\Psi_{dr} = L_3(I_{dr1} - I_{qr2}) + M_{d1r} I_{d1} - M_{d2r} I_{q2} \quad (11)$$

$$\Psi_{qr} = L_3(I_{qr1} - I_{dr2}) + M_{q1r} I_{q1} - M_{q2r} I_{d2} \quad (12)$$

where briefly, R and L – are the model's equivalent resistances and impedances (own and respectively mutual coupling); M – electromagnetic torques; Ψ – fluxes; ω – rotational speed; U and I – voltages and currents; i=1,2,3 indexes corresponding to the three machine's windings – for rotor and stator (in stator are placed the load and excitation windings).

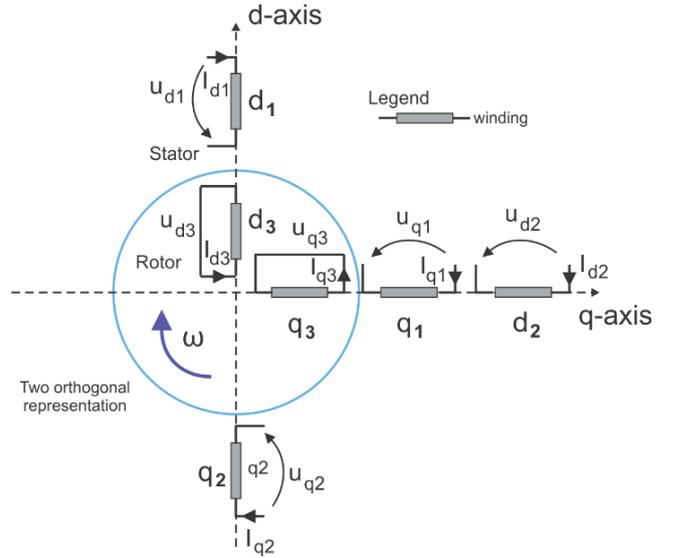

Figure 1. The double winded three-phased induction generator – two orthogonal representation.

## III. CONSIDERATIONS ON MINIMUN VARIANCE ADAPTIVE CONTROL STRUCTURE

The classical structure for an adaptive control system based on a minimum variance controller is well known in technical literature and is presented in figure 2. The structure is also detailed in other papers [3],[5],[6],[9],[10].

It has a block that implements the control law, a parameter estimator block and the controlled plant. The control law is obtained by the minimization of a criterion function that has two terms, one for assuring the minimum variance of the controlled output error and respectively the second term assuring the minimum variance for the controller output.

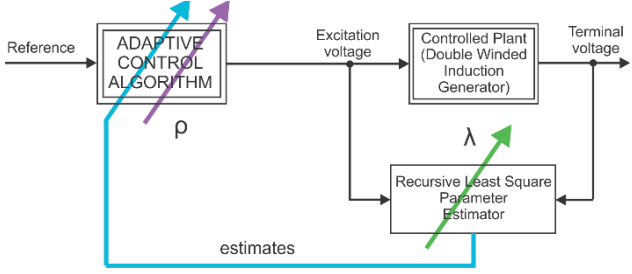

Figure 2. Adaptive control system structure.

In the paper there was considered, for the design of the control law, that the double winded induction generator can be modelled by using a 5$^{th}$ order discrete transfer function (near an operating point) (relation 13).

The considered discrete transfer function is:

$$H(q^{-1}) = \frac{y}{u} = q^{-1}\frac{B(q^{-1})}{A(q^{-1})} =$$
$$= q^{-1}\frac{b_4 q^{-4}+b_3 q^{-3}+b_2 q^{-2}+b_1 q^{-1}+b_0}{a_4 q^{-4}+a_3 q^{-3}+a_2 q^{-2}+a_1 q^{-1}+1} \quad (13)$$

The estimator block is a classical recursive least square (RLS) parameter estimator, that has as tuning parameter a forgetting factor ($\lambda$).

Using the 5$^{th}$ order linearized model of the induction generator and the criterion function there was obtained the following discrete control law (relation 14) [3]:

$$u_t = \frac{w_t - q[1-\hat{A}(q^{-1})]y_t}{\hat{B}(q^{-1})+\rho} + \frac{\rho}{\hat{B}(q^{-1})+\rho}u_t^* \quad (14)$$

where: $w_t$– the reference; $y_t$ – the controlled terminal voltage (controlled output) at discrete time $t$; $u_t$-the controller output (excitation voltage); $u_t^*$– the steady state controller output; $\rho$ – control penalty factor; $\hat{A}$ and $\hat{B}$ are the estimates of $A$ and $B$ polynomials; $q^{-1}$- the shift operator.

In order to ensure the numerical excitation of the estimator there was used a stochastic noise with variance $\sigma^2$=0.01 for the disturbing of the controlled plant [3],[7], [8].

The control structure was tested and validated for the case of the excitation control system for the induction generator with separate excitation winding.

IV. SIMULATION STUDY CASES

A. *Simulation study cases for the validation of the double winded induction generator model.*

In order to validate the proposed model for the induction generator there were performed several simulation study cases. All figures are scaled in per unit (p.u.) and the time unit is expressed in seconds.

In figure 3, is presented the relative terminal voltage variation in the case of an external perturbation (mechanical torque) with a step variation (increase with 10%) at time moment t=5 (sec).

In figure 4, is depicted the evolution of the controlled output in the case that consumer resistance increases with 10%, practically this being the case of an unload (disconnected consumer). The obtained results confirm the dynamic phenomena of the electrical machine and are comparable with the one found in technical literature.

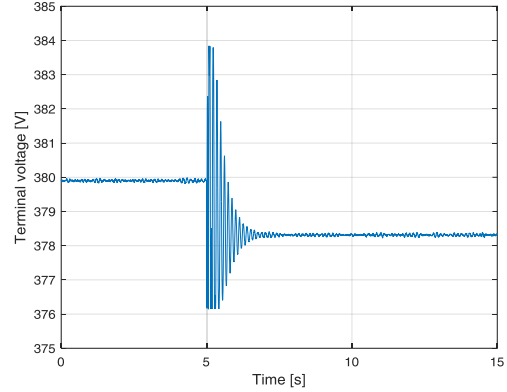

Figure 3. The terminal voltage variation – mechanical torque increased with 10%.

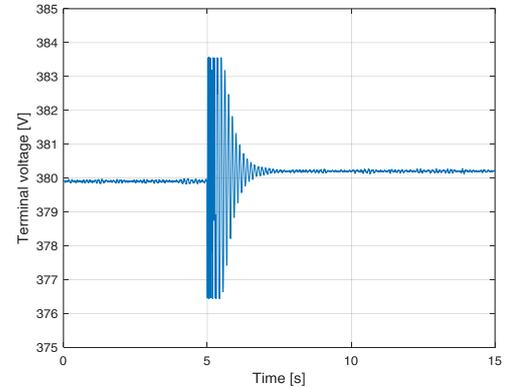

Figure 4. The terminal voltage variation – load decreased with 10% (unload).

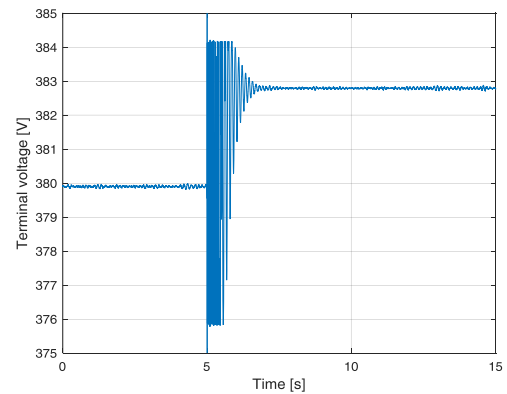

Figure 5. The terminal voltage variation – excitation voltage increased with 10%

In figure 5, is presented the terminal voltage evolution, in the case of an increase with 10% of the excitation voltage.

## B. Simulation study cases for the validation of the adaptive control structure.

In this paragraph are presented simulation study cases in order to validate the proposed adaptive control structure of $5^{th}$ order.

In table 1, for an easier understanding, there were synthetically presented the most significant simulation study cases, together with their corresponding pair of values for the forgetting factor ($\lambda$) and respectively the control penalty factor ($\rho$).

TABLE I. Synthetic table for the performed simulations.

| Case study number | Conditions | | Brief comments |
|---|---|---|---|
| | $\lambda$ – forgetting factor | $\rho$ - control penalty factor | |
| I | 0.995 | 0.0750 | Acceptable results |
| II | 0.995 | 0.0720 | Poor results |
| III | 0.995 | 0.0725 | Best results |
| IV | 0.995 | 0.0730 | Good results |
| V | 0.990 | 0.0725 | Poor results |

The performed simulations have as objective to establish the most suitable pair of two parameter ($\lambda$, $\rho$) from the point of view of the obtained results (control, controller's output adequate penalty). There were performed may simulation study cases under different conditions. In the paper are presented only the most relevant cases in order to validate the proposed adaptive control structure of $5^{th}$ order. Also, in the paper are presented only the simulations that considered the variation of the mechanical torque with an increase of 10%.

In the simulation there were considered several values for the pair of two parameter ($\lambda$, $\rho$), but the best results for forgetting factor of the RLS estimator was $\lambda$=0.995. Therefore, the presented cases mainly consider this value for forgetting factor (all study cases, except study case II).

The reason for considering only the mechanical torque variation (that acts as an external perturbation) is that double winded induction generator is usually part of wind energy conversion systems (WECS). In the case of WECS the mechanical torque variations are normal phenomena as they occur due to frequent wind speed changes.

*Study case **I**.*

In this study case the obtained results are acceptable, but still not enough good from the control's point of view (overshoots, oscillations, settling time). As it can be seen in figure 6.a, settling time is up to 7 seconds, but overshoots are acceptable (only 5[V]). In figure 6.b, is presented the excitation voltage evolution, where the voltage spikes are lower (maximum 485 [V]).

The captions presented in paper have the abscise axis scaled in time units (seconds) and on the ordinate axis is presented the analyzed variable (mainly in [V], excepting the parameter estimates, where the parameters are of course dimensionless.

In figure 6.c, are represented the parameter estimates evolutions. In this case has been considered a forgetting factor $\lambda = 0.995$ (meaning a higher memory of the RLS estimator), fact that leads to the represented evolution of the A and B polynomials.

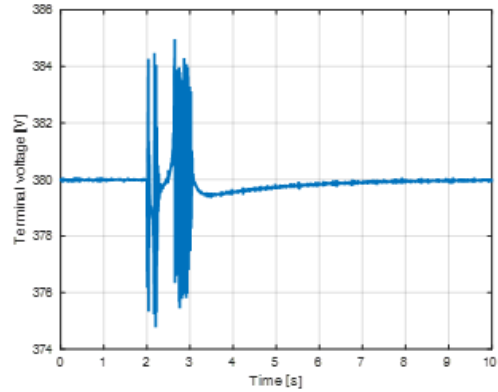

Figure 6.a. Terminal voltage evolution.

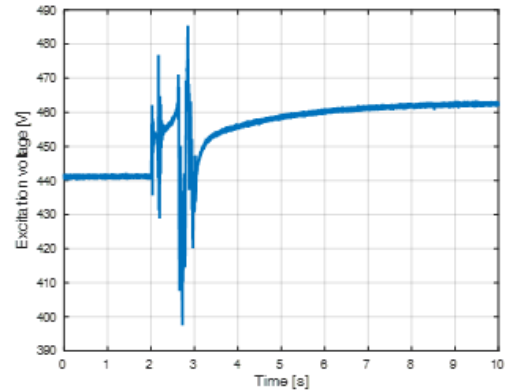

Figure 6.b. Excitation voltage evolution.

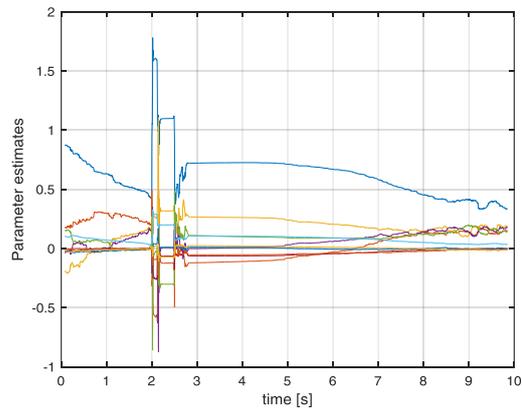

Figure 6.c. Parameter estimates evolution.

*Study case **II**.*

The results obtained in this first study case are quite poor. In figure 7.a, there can be noticed a quite long settling time (5 sec) and an overshoot up to 13 [V]. In figure 7.b is presented the evolution of the excitation voltage, where can be noticed quite high spikes of voltage (up to 540 [V]).

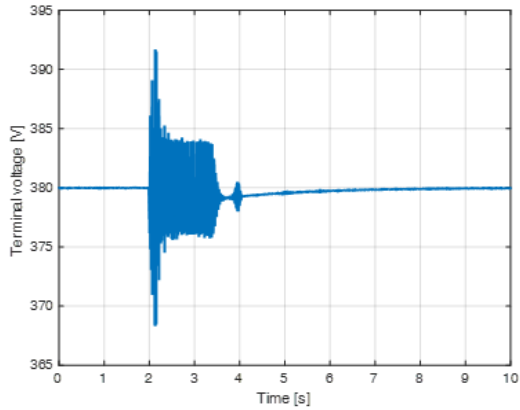

Figure 7.a. Terminal voltage evolution.

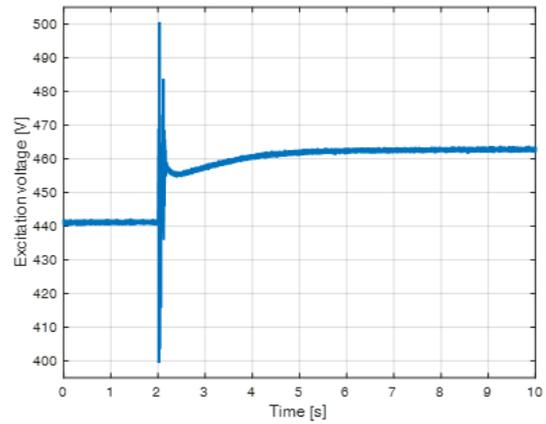

Figure 8.b. Excitation voltage evolution.

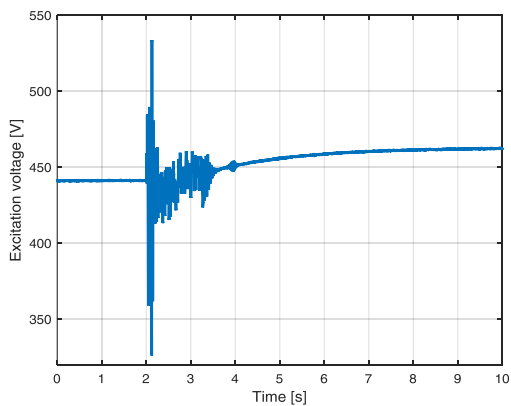

Figure 7.b. Excitation voltage evolution.

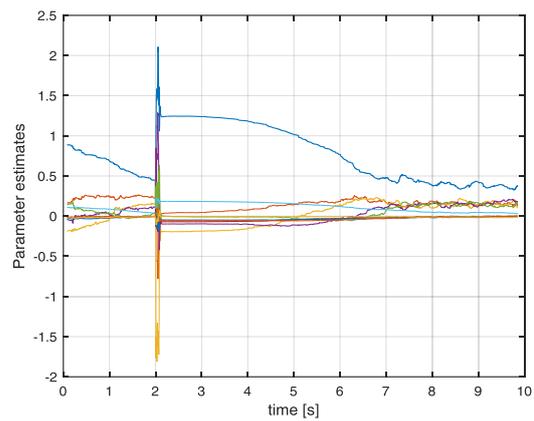

Figure 8.c. Parameter estimates evolution.

*Study case **III***.

In this study case can be observed the best results, corresponding to the pair of values ($\lambda$=0.995 and $\rho$=0.0725). In figure 8.a, is depicted the terminal voltage evolution. There can be noticed a short settling time (only 2 seconds) and a quite small overshoot of 8 [V] only. In figure 8.b is presented the excitation voltage evolution with settles at 462 [V]. Also, in figure 8.c are shown the parameter estimates evolutions (practically the A and B polynomials).

*Study case **IV***.

In this study case can be noticed good results. In figure 9.a, is being represented the terminal voltage evolution (the controlled output), the settling time being only 3 seconds, the overshoot is quite small (8 [V]).

In figure 9.b, the is represented the excitation voltage evolution (the controller's output), with spikes up to 495 [V].

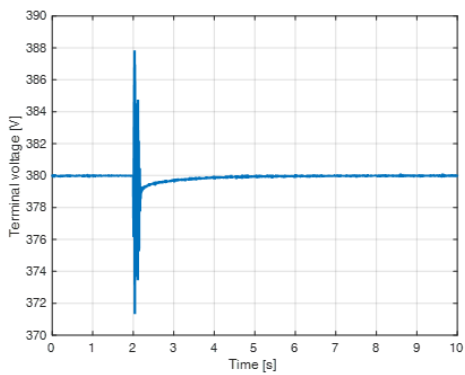

Figure 8.a. Terminal voltage evolution.

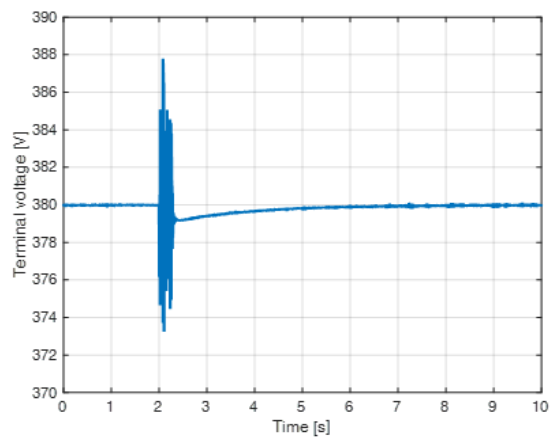

Figure 9.a. Terminal voltage evolution.

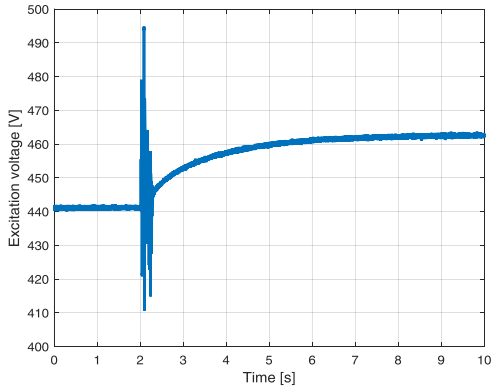

Figure 9.b. Excitation voltage evolution.

*Study case V.*

The results in this study case are also unsatisfactory, in figure 10.a, there can be observed overshoot of up to 12 [V] amplitude and significant oscillations. The settling time is quite long (5 – 6 sec). In figure 10.b, is presented the excitation voltage evolution (the controller's output). Also, in this case occur high spikes - up to 550 [V].

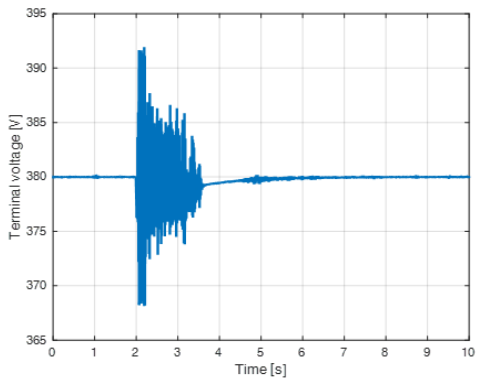

Figure 10.a. Terminal voltage evolution.

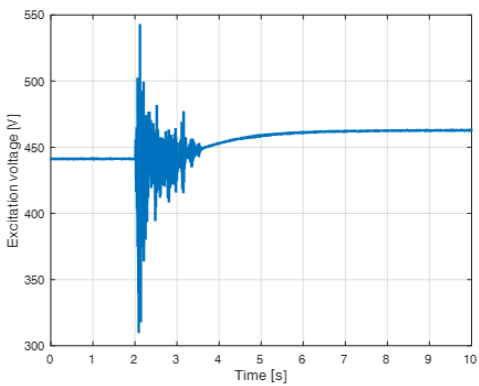

Figure 10.b. Excitation voltage evolution.

The obtained results prove the validity of the proposed adaptive control structure based on a linearized model of $5^{th}$ order. However, its complexity is not quite justified because the authors have proven in previous paper that a similar adaptive control structure based on $4^{th}$ order generator model has very good results also, maybe even better, in the case of consumer load/unloads, mechanical torque increases/ decreases and so on [3]. There can be drawn the conclusion that a higher order adaptive control structure doesn't necessarily mean implicit better results (there occurs a trade off regarding its higher complexity and computational requirements).

## V. CONCLUSIONS

The considered control structure is an adaptive type and was tested and validated for the excitation control for a double winded induction generator connected to a power system and consumers. The classical structure of a minimum variance control system, designed by minimization of a criterion function, is modified in order to solve the problems made by usual external perturbations (such as: variation of the mechanic torque, consumers load/unload).

There are performed simulation study cases that prove the validity of reduced order model ($5^{th}$ order) for the induction machine, used only to design a simplified control law. Also, there is validated the proposed control structure in the considered operating regimes.